\documentclass[12pt]{iopart}
\usepackage{iopams}
  
\usepackage{epsfig}  
\newcommand{\beq}{\begin{equation}}
\newcommand{\eeq}{\end{equation}}
\newcommand{\bea}{\vspace{0.25cm}\begin{eqnarray}}
\newcommand{\eea}{\end{eqnarray}}

\newcommand{\ro}{\mbox{{\boldmath$\rho$}}}
\newcommand{\qb}{\mbox{{\bf q}}}
\newcommand{\pb}{{{\bf p}}}
\newcommand{\bb}{{{\bf b}}}

\def\lsim{\mathrel{\rlap{\lower4pt\hbox{\hskip1pt$\sim$}}
    \raise1pt\hbox{$<$}}}         
\def\gsim{\mathrel{\rlap{\lower4pt\hbox{\hskip1pt$\sim$}}
    \raise1pt\hbox{$>$}}}         

\begin{document}

\title[]{Nuclear suppression  of light hadrons and
single electrons at RHIC and LHC
}

\author{B.G. Zakharov}

\address{
 L.D.~Landau Institute for Theoretical Physics,
        GSP-1, 117940,\\ Kosygina Str. 2, 117334 Moscow, Russia
}
\ead{bgz@itp.ac.ru}
\begin{abstract}
We examine  whether it is possible 
to simultaneously describe  the experimental 
data from RHIC and LHC on nuclear suppression of light hadrons and
non-photonic single electrons 
in the pQCD picture of parton
energy loss. 
We perform calculations accounting for  
both radiative and collisional energy loss.
We  show that once the coupling constant 
is fixed from comparison with data on the nuclear 
modification factor for light
hadrons it gives a satisfactory agreement with 
data on the electron $R_{AA}$ and azimuthal anisotropy $v_2$.
Our results show that the collisional mechanism 
is only of marginal significance in nuclear suppression
of single electrons
except for the bottom contribution 
at momenta $\lsim 6 - 8$ GeV. 

\end{abstract}

\maketitle

\section{Introduction}
A remarkable result of the experiments on $AA$-collisions at RHIC 
\cite{PHENIX1_e,STAR_e,PHENIX2_e} and LHC \cite{ALICE_e} is
the observation that nuclear suppression of high-$p_{T}$ single 
electrons from semi-leptonic decays of heavy mesons is almost 
as strong as that of pions. 
It is believed that nuclear suppression  of high-$p_{T}$
particles (jet quenching)  is due to 
radiative \cite{GW,BDMPS,LCPI,W1,GLV1,AMY} and collisional 
\cite{Bjorken1} parton energy loss
in the hot quark-gluon plasma (QGP) produced in the initial stage of 
$AA$-collisions.
The observed  strong suppression of single electrons
indicates that 
heavy quarks are quenched approximately as light
ones. 
This seemed to be somewhat puzzling in light of the prediction of the 
dead cone reduction of the heavy quark radiative energy loss \cite{DK}.
In recent years nuclear suppression of non-photonic single electrons
due to radiative and collisional heavy quark energy loss
in pQCD
received considerable theoretical attention  
\cite{ASW_e,MD1,WHDG_e1,WG,Aich,MD2,Greiner1,Greiner2}.
However, no clear consensus has emerged on 
the physical picture of the effect and whether the pQCD 
can explain it.

Comparison of the radiative
energy loss (calculated within the light-cone path integral 
(LCPI) approach \cite{LCPI,BSZ}) and the collisional energy loss 
calculated with the same $\alpha_{s}$ and 
the Debye screening mass shows \cite{Z_Ecoll}  that
for quark energy $E\gsim 5$ GeV the collisional energy loss 
is relatively small for light quarks and $c$-quark, and 
for $b$-quark the collisional energy loss becomes important
only at $E\lsim 10$ GeV.
This shows that the collisional energy loss 
should be of only marginal significance in suppression of single electrons
at $p_{T}\gsim 5$ GeV (since the 
electron spectrum is controlled by heavy quark production at approximately 
twice the electron momentum). 
This is also supported from 
computations of the nuclear modification factor $R_{AA}$ of single electrons
for purely collisional
mechanism in \cite{Greiner1,Greiner2},
where it was found that to fit the data  the cross sections 
of the $2\to 2$ processes should be enhanced by a factor $\sim 4$.
Thus, it is natural to expect that, if the pQCD is 
valid for parton energy loss for RHIC and LHC conditions,
the nuclear suppression of single electrons  should be described in
a picture where the radiative mechanism dominates.

The purpose of the present work is to analyze the available RHIC and LHC data
on the electron suppression, and to examine whether the observed
nuclear suppression of single electrons and light hadrons  can be described 
simultaneously in the pQCD.
In our study we use 
the LCPI approach \cite{LCPI} to induced gluon emission.
The advantage of this formalism is that it treats accurately 
the mass and finite-size effects, and is valid beyond the soft
gluon approximation (used in previous analyses \cite{ASW_e,WHDG_e1,WG,MD2}).
Calculations beyond the soft gluon 
approximation are especially desirable for $c$-quark.
Indeed, in the LCPI formalism \cite{LCPI} the induced gluon
$x$-spectrum ($x$ is the gluon fractional momentum)
is expressed through the solution of 
a two-dimensional Schr\"odinger equation in which the
longitudinal coordinate $z$ (along the fast parton momentum)
plays the role of time. And the Hamiltonian (see Appendix) depends 
on the parton 
masses only through the term 
$1/L_{f}=[m_{q}^{2}x^{2}+m_{g}^{2}(1-x)^{2}]/2x(1-x)E$,  where
$E$ is the initial quark energy, 
$m_{q,g}$ are the quasiparticle parton masses.
The quark mass becomes important when 
$m_{q}^{2}x^{2}\gsim m_{g}^{2}(1-x)^{2}$. 
Taking $m_{g}\sim 400$ MeV \cite{LH}, we see 
that it occurs 
at $x\gsim 0.3$ for $c$-quark and $x\gsim 0.1$ for $b$-quark,
and below these values the sensitivity to the quark mass
should vanish quickly.
Accurate numerical calculations \cite{AZ} corroborate these qualitative
estimates. Note that the results of \cite{AZ}
show that 
at energy $\sim 10 - 20$ GeV
for a finite-size plasma the quark mass suppression of the 
radiative energy loss may be considerably weaker than 
predicted in the dead cone model \cite{DK}, and at higher
energies the radiative energy loss may even be enhanced for 
heavy quarks.
Therefore the observed strong suppression of single
electrons does not seem to be very strange.

We treat the effect of parton energy loss on 
the single electron yield within  
the scheme developed previously for light hadrons \cite{RAA08} (see
also \cite{Z_RHIC-ALICE}). It takes into account both radiative and
collisional energy loss, and fluctuations of the fast parton
path lengths in the QGP. The calculations of radiative and 
collisional energy loss are performed with
running coupling. 
In our recent note \cite{RAA12} we analyzed within this approach the 
flavor dependence of $R_{AA}$ at the LHC energy.
In this paper we concentrate on suppression of single electrons
and analyze both the RHIC and LHC data. Besides the nuclear modification
factor $R_{AA}$ we present results for the azimuthal asymmetry $v_{2}$.

\section{Main features of the model}
In this section
we present the basic features of our approach. 
We refer the interested reader to Refs. \cite{RAA08,Z_RHIC-ALICE} for 
more details.

We define the nuclear modification factor $R_{AA}$ 
for a given impact parameter $b$ as
\beq
R_{AA}(b,\pb_{T},y)=\frac{{dN(A+A\rightarrow h+X)}/{d\pb_{T}dy}}
{T_{AA}(b){d\sigma(N+N\rightarrow h+X)}/{d\pb_{T}dy}}\,,
\label{eq:10}
\eeq
where $\pb_{T}$ is the particle transverse momentum, $y$ is rapidity (we
consider the central region around $y=0$), 
$T_{AA}(b)=\int d\ro T_{A}(\ro) T_{A}(\ro-\bb)$,
$T_{A}$ is the nucleus profile function.
The numerator in (\ref{eq:10}) is the differential yield of the 
process $A+A\to h+X$
(for clarity we omit the argument $b$) given by
\beq
\frac{dN(A+A\rightarrow h+X)}{d\pb_{T} dy}=\int d\ro T_{A}(\ro)T_{A}(\ro-\bb)
\frac{d\sigma_{m}(N+N\rightarrow h+X)}{d\pb_{T} dy}\,,\,\,\,
\label{eq:20}
\eeq
where 
$d\sigma_{m}/d\pb_{T} dy$
is the medium-modified hard cross section.
Similarly to the ordinary pQCD formula we write it as
\beq
\frac{d\sigma_{m}(N+N\rightarrow h+X)}{d\pb_{T} dy}=
\sum_{i}\int_{0}^{1} \frac{dz}{z^{2}}
D_{h/i}^{m}(z, Q)
\frac{d\sigma(N+N\rightarrow i+X)}{d\pb_{T}^{i} dy}\,,\,\,\,
\label{eq:30}
\eeq
where $\pb_{T}^{i}=\pb_{T}/z$ is the transverse momentum
of the initial hard parton, 
${d\sigma(N+N\rightarrow i+X)}/{d\pb_{T}^{i} dy}$ is the
ordinary hard cross section,  
$D_{h/i}^{m}$ is the medium-modified fragmentation function (FF)
for transition of a parton $i$ into the observed particle $h$.
For the initial virtuality $Q$ we use the parton momentum $p^{i}_{T}$.
As in \cite{RAA08}, the hard cross sections on the right-hand side of
(\ref{eq:30})  were calculated using the LO 
pQCD formula with the CTEQ6 \cite{CTEQ6} parton distribution functions.
The higher order effects were simulated taking 
for the virtuality scale in $\alpha_{s}$ the value 
$cQ$ with $c=0.265$ as in the PYTHIA event generator \cite{PYTHIA}.
This prescription gives a fairly good description 
of the $p_{T}$-dependence of the spectra in $pp$-collisions
\footnote{For the heavy quark
cross sections our LO formulas give the $p_{T}$-dependences 
(and the $c/b$ ratio) that agree well with the more sophisticated 
FONLL calculations \cite{Vogt}. However, the normalization of the cross
sections are smaller by a factor $\sim 1/2$. But for $R_{AA}$ 
it is not important.}.
We account for 
the nuclear modification of the parton densities
with the 
EKS98 correction \cite{EKS98}
(which gives a small deviation of $R_{AA}$ from unity even without
parton energy loss).

The formation length arguments allow, in first approximation,
to neglect the overlap between the DGLAP and the induced stages
of the parton showering \cite{RAA08}. Then, assuming that formation
of the final particle $h$ occurs outside the medium, 
symbolically the medium-modified FF can be written as
\beq
D_{h/i}^{m}(Q)\approx D_{h/j}(Q_{0})
\otimes D_{j/k}^{in}\otimes D_{k/i}(Q)\,,
\label{eq:40}
\eeq
where $\otimes$ denotes $z$-convolution, 
$D_{k/i}$ is the ordinary DGLAP FF for $i\to k$ parton transition,
$D_{j/k}^{in}$ is the FF for $j\to k$ parton transition in the QGP
due to induced gluon emission, and 
$D_{h/j}$ describes fragmentation of the parton $j$ into
the detected particle $h$ outside of the QGP.

We computed the DGLAP FFs using the PYTHIA event 
generator \cite{PYTHIA}.
For the stage outside the QGP
for light partons
we use for the $D_{h/j}(Q_{0})$ the 
KKP \cite{KKP} FFs  with $Q_{0}=2$ GeV.
We treat the formation of single electrons
from  heavy quarks as the two-step fragmentations
$c\to D\to e$ and $b\to B\to e$. 
For the $c\to D$ and $b\to B$ transitions we use 
the Peterson FF with parameters
$\epsilon_{c}=0.06$ and $\epsilon_{b}=0.006$.
The $z$-distribution for the $M\to e$ transitions (for $M=B/D$)
that we need may be expressed via the electron 
momentum spectrum $dB/dp$ in the heavy meson rest frame as
\beq
D_{e/M}(z,P)=\frac{P}{4}
\int_{0}^{\infty}
dq^{2}\frac{\cosh(\phi-\theta)}
{p^{2}\cosh{\phi}}
\cdot\frac{dB}{dp}\,,
\label{eq:50}
\eeq
where $p=\sqrt{(q^{2}+m_{e}^{2})\cosh^{2}(\phi-\theta)-m_{e}^{2}}$,
$\theta=\mbox{arcsinh}(P/M)$, $\phi=\mbox{arcsinh}(zP/\sqrt{q^{2}+m_{e}^{2}})$,
$P$ is the heavy meson momentum, and $M$ is its mass.  
We evaluated the $B/D\to e$  FFs 
using the CLEO data 
\cite{CLEO_B,CLEO_D} on the electron spectra in the $B/D$-meson decays.
We did not include the $B\to D\to e$ process, which gives a negligible 
contribution \cite{Vogt}.

To calculate the FFs $D_{j/k}^{in}$ in the induced stage we
use the one gluon spectrum $dP/dx$ computed in the LCPI formalism \cite{LCPI}
with the help of the method suggested in
\cite{Z04_RAA}. The formulas for calculation of $dP/dx$ are recorded
in the Appendix for the reader's convenience. The effect of 
multiple gluon emission was accounted for 
using Landau's method as in \cite{BDMS_RAA} (see \cite{RAA08} for
details).
As in \cite{RAA08}, for the  
quasiparticle masses of light quarks and gluon 
we take $m_{q}=300$ and $m_{g}=400$ MeV  supported by 
the analysis of the lattice data \cite{LH}. The results are practically
insensitive to the light quark mass.
For heavy quarks we take $m_{c}=1.2$ GeV
and $m_{b}=4.75$ GeV. We use 
the Debye mass obtained in the lattice calculations \cite{Bielefeld_Md} 
giving $\mu_{D}/T$ slowly decreasing with $T$  
($\mu_{D}/T\approx 3$ at $T\sim 1.5T_{c}$, $\mu_{D}/T\approx 2.4$ at 
$T\sim 4T_{c}$). 
We used running $\alpha_s$
frozen at some value $\alpha_{s}^{fr}$ at low momenta
(the technical details for incorporating the running $\alpha_s$
can be found in \cite{Z04_RAA}). For gluon emission
in vacuum a reasonable choice is $\alpha_{s}^{fr}\approx 0.7$
\cite{NZ_HERA,DKT}. However, in the QGP the thermal effects can suppress
the $\alpha_{s}^{fr}$, and we regard it as a free parameter
which should be fixed by the data.
If our model is valid, the $\alpha_{s}^{fr}$ for
light hadrons and single electrons should be close to each other.

As in \cite{RAA08},
we treat the collisional mechanism as a perturbation to the radiative 
one. 
We account for its effect by redefining the initial QGP 
temperature in calculating the radiative medium-modified FFs according to the
condition
\beq
\Delta E_{rad}(T^{\,'}_{0})=\Delta E_{rad}(T_{0})+\Delta E_{col}(T_{0})\,,
\label{eq:60}
\eeq
where
$\Delta E_{rad/col}$ is the radiative/collisional energy loss, $T_{0}$
is the real initial temperature of the QGP, and $T^{\,'}_{0}$ is the 
renormalized temperature. 
We carry out this temperature renormalization 
for each parton trajectory in the QGP
(separately for quarks and gluons).
For the collisional energy loss we use the 
Bjorken method \cite{Bjorken1}
with an accurate treatment of kinematics of the $2\to 2$ 
processes (the details can be found
in \cite{Z_Ecoll}). For the collisional mechanism we use the same 
parametrization of $\alpha_{s}(Q)$ as for the radiative one.
Both the radiative and collisional contributions in (\ref{eq:60})
were calculated for maximum energy transfer constrained  by half of the 
initial parton energy.

\section{Numerical results and comparison with the data}
We have performed the computations using 
Bjorken's 1+1D expansion of the QGP \cite{Bjorken2}, 
which gives $T_{0}^{3}\tau_{0}=T^{3}\tau$. We take $\tau_{0}=0.5$ fm.
In calculating the medium-modified FFs, for simplicity, we neglect 
variation of the initial temperature $T_{0}$ with the 
transverse coordinates across the overlapping area of two colliding
nuclei. We define this area as overlapping of two 
circles with radius $R=R_{A}+k d$, where $R_{A}$ and $d$
are the parameters of the Woods-Saxon nuclear
density $\rho_{A}(r)=\rho_{0}/[1+\exp((r-R_{A})/d)]$. We take 
$k=1.5$, which guarantees that the fraction of the lost
QGP volume is negligible. The results are not very sensitive 
to variation of $k$ in the physically reasonable range 
$1\lsim k\lsim 2$ (except for very peripheral $AA$-collisions,
which we do not address in the present paper).
To fix $T_{0}$ (in each centrality bin) we use data on the charged 
hadron multiplicity pseudorapidity density $dN_{ch}/d\eta$ 
from RHIC \cite{STAR_Nch} and LHC \cite{CMS_Nch,ALICE_Nch}.
For the entropy/multiplicity ratio we use
$dS/dy{\Big/}dN_{ch}/d\eta\approx 7.67$ obtained in \cite{BM-entropy}.
For the chemically equilibrated ideal QGP (we take $N_{f}=2.5$) it gives 
$T_{0}\approx 320$ MeV for central
Au+Au collisions at $\sqrt{s}=200$ GeV, and
$T_{0}\approx 420$ MeV for central
Pb+Pb collisions at $\sqrt{s}=2.76$ TeV.
For each hard parton we calculate accurately 
the path length in the QGP, $L$, according to the geometry
of the $AA$-collision.
Since  the QGP should cool quickly 
at times about $1 - 2$ units of 
the nucleus radius due to transverse expansion
\cite{Bjorken2}, we  
impose the condition $L< L_{max}$. We performed the computations for 
$L_{max}=8$. We checked that the bigger value  $L_{max}=10$ fm gives 
almost the same.
\begin{figure} [t]
\begin{center}
\epsfig{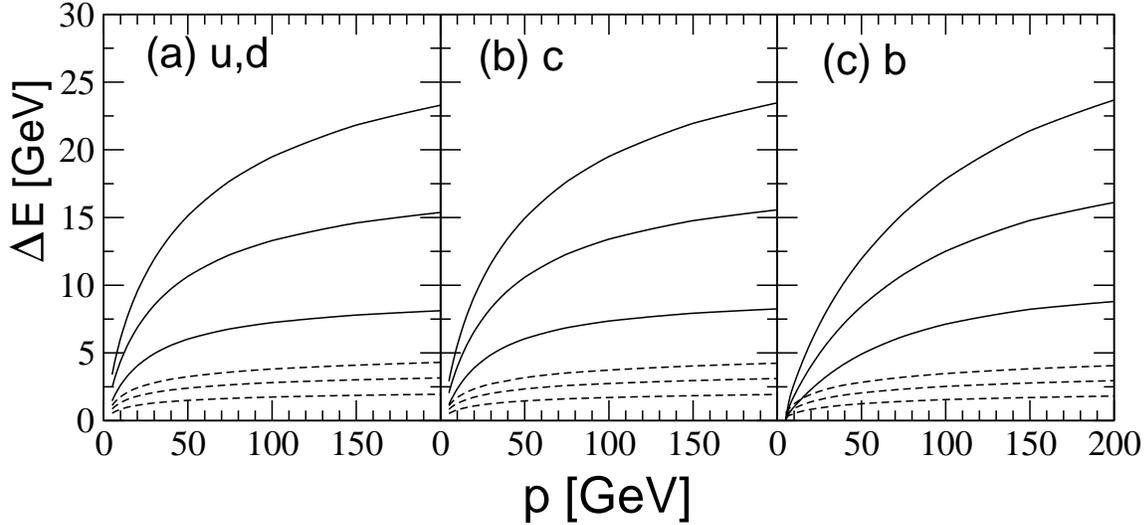}
\end{center}
\caption[.]{
The radiative (solid) and collisional (dashed) energy loss
for (a) $u,d$, (b) $c$, and (c) $b$ quarks in expanding plasma
of size $L=5$ fm for $T_{0}=300$, $400$, and $500$ MeV
for $\alpha_{s}^{fr}=0.5$. The order of the curves corresponds to
ordering of their $T_{0}$.}
\end{figure}

In order to illustrate the relative contribution of the collisional
mechanism to parton energy loss  in our model, 
in Fig.~1 we present $\Delta E_{rad}$ and $\Delta E_{col}$ 
for light and heavy quarks at
$T_{0}=$300, 400 and 500 MeV obtained 
for $L=5$ fm, which is a typical parton path length for central $AA$-collisions
at RHIC and LHC. One can see that for light quarks and $c$-quarks
the contribution of the collisional mechanism is relatively
small at $p\gsim 5$ GeV. But for $b$-quarks the collisional mechanism
becomes clearly important at $p\lsim 10$ GeV. In this region our prescription
(\ref{eq:60}) does not apply. 
From the variation of $\Delta E$ for the radiative and collisional mechanisms
with the initial plasma temperature in Fig.~1 one can understand 
the magnitude of the ratio $T_{0}^{'}/T_{0}$
at $T_{0}\sim 300 - 400$ MeV relevant to RHIC and LHC. For
light quarks and $c$ quark  at $p\sim 10 - 50$ GeV 
$(T_{0}^{'}/T_{0})^{3}\sim 1.3 - 1.5$,
and for $b$ quark at $p\sim 10 - 20$ GeV this ratio is somewhat larger 
$\sim 1.5 - 1.8$.

\begin{figure} [t]
\begin{center}
\epsfig{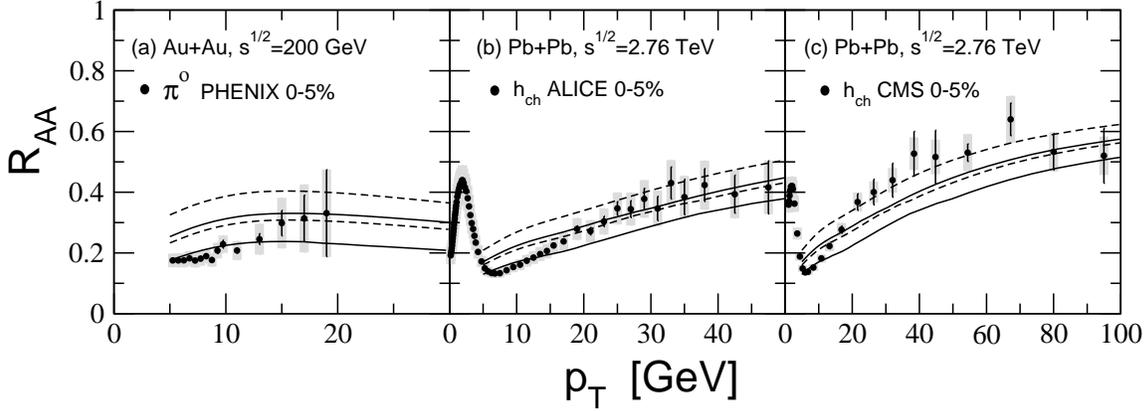}
\end{center}
\caption[.]
{
(a) $R_{AA}$ for $\pi^{0}$ for 0-5\% central Au+Au collisions
at $\sqrt{s}=200$ GeV from our calculations compared to data from 
PHENIX \cite{PHENIX_RAA_pi}. 
(b,c) $R_{AA}$ for charged hadrons for 0-5\% central Pb+Pb collisions
at $\sqrt{s}=2.76$ GeV from our calculations compared to data from 
(b) ALICE \cite{ALICE_RAAch} and (c) CMS \cite{CMS_RAAch}.
Systematic experimental errors are shown as shaded areas. 
The curves show our calculations for radiative and collisional
energy loss (solid), and for purely radiative energy loss (dashed)
for $\alpha_{s}^{fr}=0.4$
(upper curves) and 0.5 (lower curves).
}
\end{figure}

Fig.~2 shows comparison of our predictions for $R_{AA}$ for 
0--5\% centrality bin
for (a) $\pi^{0}$-meson in  
Au+Au collisions at $\sqrt{s}=200$ GeV
to PHENIX data  \cite{PHENIX_RAA_pi},
and for (b,c) charged hadrons in 
Pb+Pb collisions at $\sqrt{s}=2.76$ TeV
to (b) ALICE \cite{ALICE_RAAch} and (c) CMS \cite{CMS_RAAch} 
data. Note that in our calculations based on (\ref{eq:40}) we 
ignore possible anomalous baryon contribution \cite{AZ_baryon} 
to the yield of charged particles. Theoretically it is expected
to be small at LHC \cite{AZ_baryon}. In our model with the KKP FFs \cite{KKP}
for the hadronization outside the QGP, $R_{AA}$ for charged hadrons
turns out to be very close to that for pions. Experimentally the 
preliminary data from ALICE \cite{ALICE-pi0} on $R_{AA}$ for neutral pions
also corroborate this.
We present our results for $\alpha_{s}^{fr}=0.4$ (upper curves) and 
0.5 (lower curves).
To illustrate the effect of collisional energy loss
we show the total $R_{AA}$ with radiative and collisional energy loss
(solid) and for purely radiative energy loss
(dashed).
One can see that the effect of the collisional mechanism
is relatively small (especially for LHC).
We present the results for $p_{T}\gsim 5$ GeV since for  smaller momenta 
our calculations of the induced gluon emission (based on the 
relativistic approximation) are hardly robust.
Fig.~2 shows that for light hadrons 
the window $\alpha_{s}^{fr}\sim 0.4 - 0.5$ 
leads to a reasonable magnitude of $R_{AA}$.
However, the agreement in the shape of the $p_{T}$-dependence
of $R_{AA}$ is evidently not perfect.
But this discrepancy does not seem to be very dramatic
since the theoretical uncertainties of the approximations 
involved may be significant.
One of the most serious sources of the theoretical errors, 
that can be important for the
$p_{T}$-dependence of $R_{AA}$, is the Landau
approximation for  multiple gluon emission \cite{BDMS_RAA}.
As can be seen from Fig.~2a for RHIC the agreement 
of the theoretical $R_{AA}$
(radiative plus collisional energy loss) with the data  
is better for $\alpha_{s}^{fr}=0.5$. Figs.~2b,c show that for LHC 
the value $\alpha_{s}^{fr}=0.4$ seems to be preferred by the data
(if one considers the complete $p_{T}$ range).
Thus, the values $\alpha_{s}^{fr}=0.5$ and $0.4$ seem to 
be reasonable benchmarks for calculations of the nuclear modification factor
of single electrons at RHIC and LHC energies.
The tendency of the decrease of $\alpha_{s}^{fr}$ from RHIC to LHC, 
first observed in \cite{Z_RHIC-ALICE}, is natural, since 
the thermal reduction of $\alpha_{s}$ should be stronger at the LHC energies.
In fact, the variation of $\alpha_{s}^{fr}$ may be stronger 
if one takes $\tau_{0}^{RHIC}>\tau_{0}^{LHC}$, 
which seems to be quite reasonable since from the dimension arguments
one can expect $\tau_{0}\propto 1/T_{0}$.
However, our purpose is to study the variation of nuclear suppression
from light flavors to heavy ones probed via single electrons, and for
this reason it is sufficient to have just $\alpha_{s}^{fr}$ fixed at 
each energy from the light hadron data.
\begin{figure} [t]
\begin{center}
\epsfig{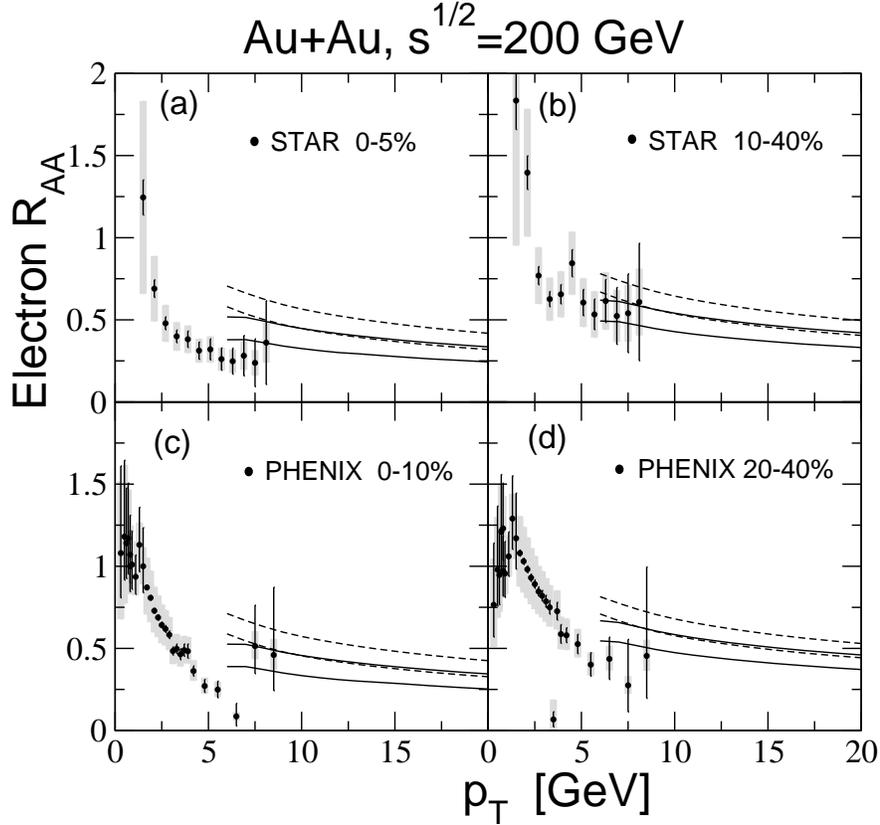}
\end{center}
\caption[.]
{
The electron $R_{AA}$ in Au+Au collisions at $\sqrt{s}=200$ 
GeV for (a) 0--5\%, (b) 10--40\%, (c) 0--10\%, (d) 20--40\% centrality
classes.  
The curves show calculations for radiative and collisional energy 
loss (solid), and for purely radiative energy loss (dashed) including
charm and bottom contributions for $\alpha_{s}^{fr}=0.4$
(upper curves) and 0.5 (lower curves).
Data points are from STAR \cite{STAR_e} and PHENIX \cite{PHENIX2_e}.
Systematic errors
are shown as shaded areas. 
}
\end{figure}

In Fig.~3 we compare results of our model
with STAR \cite{STAR_e} and PHENIX \cite{PHENIX2_e}
data on the electron $R_{AA}$.
Comparison to the data from ALICE  \cite{ALICE_e}
is shown in Fig.~4a.
In Fig.~3, 4a we show the total (charm plus bottom) $R_{AA}$ with (solid) 
and without (dashed) collisional energy loss.
From Figs.~3,~4a one sees that our pQCD model
for the same window of $\alpha_{s}^{fr}$ as for light hadrons
leads to quite satisfactory agreement with data on the electron $R_{AA}$.
Similarly to $R_{AA}$ for light hadrons the electron data
support $\alpha_{s}^{fr}\approx 0.5$ for RHIC,
and $\alpha_{s}^{fr}\approx 0.4$ for LHC.
Thus, the simultaneous 
description of the nuclear suppression of light hadrons and single electrons
in the pQCD picture seems quite 
possible. 
Of course, this should be taken with a caution
since the overlapping of the 
$p_{T}$ region, where our approximations make sense, 
with that studied experimentally is still rather narrow (especially for 
RHIC), and namely in this region the experimental errors are very large.
\begin{figure} [t]
\begin{center}
\epsfig{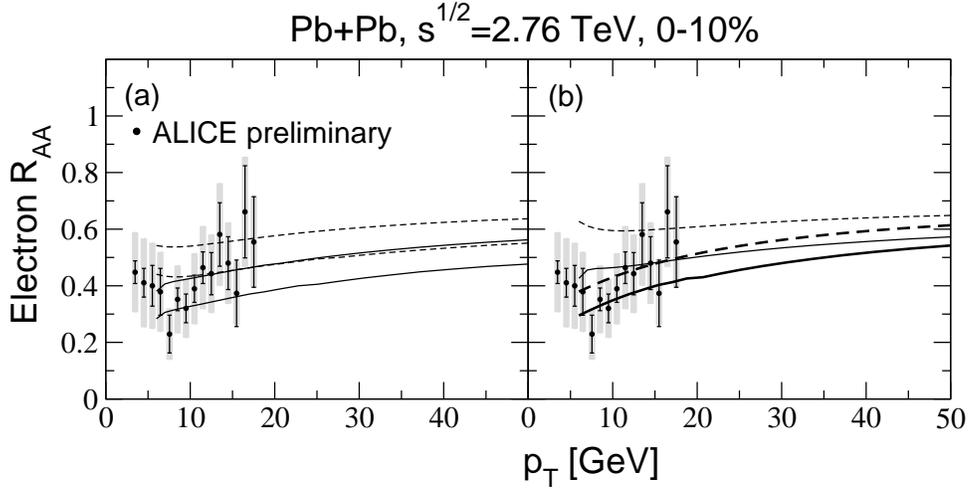}
\end{center}
\caption[.]
{
(a) The electron $R_{AA}$
for 0-10\% central 
Pb+Pb collisions at $\sqrt{s}=2.76$ TeV for $\alpha_{s}^{fr}=0.4$
(upper curves) and 0.5 (lower curves) including charm and bottom.
The experiment points are the preliminary ALICE data \cite{ALICE_e}.
Systematic errors
are shown as shaded areas.
(b) The electron $R_{AA}$ for charm (thick) and bottom (thin)
contributions at $\alpha_{s}^{fr}=0.4$.
In (a,b) the solid curves show results for radiative plus collisional
energy loss, and the dashed curves for purely radiative energy loss.
}
\end{figure}

To illustrate the effect of collisional energy loss
for $c\to e$ and $b\to e$ processes in Fig.~4b we show 
the curves for $R_{AA}$ separately for
charm (thick) and bottom (thin) with (solid) 
and without (dashed) collisional energy loss for $\alpha_{s}^{fr}=0.4$.
One sees that the collisional mechanism is more important
for bottom. The effect becomes especially significant at low $p_{T}$.
At $p_{T}\lsim 5 -  6$ GeV our treatment of the collisional mechanism 
as a perturbation to the radiative one, 
with the help of (\ref{eq:60}), loses accuracy.
Evidently, in this regime the radiative and collisional mechanisms
must be treated on an equal footing. However, a solution of this challenging
problem is still lacking.
For charm the situation is better, since 
across the whole energy range, where the relativistic
approximation makes sense, the collisional
energy loss remains relatively small.
Note that, as one can see from Fig.~4b, the difference between the 
charm and bottom suppression factors at $p_{T}\gsim 6$ GeV is relatively
small. For this reason the total (charm plus bottom) $R_{AA}$
is quite stable against variation of the $b/c$ ratio which is not very robust.
Even at $p_{T}\sim 6 - 10$ GeV variation in $b/c$ ratio of $\pm 30$\%
changes the total $R_{AA}$  by $\pm (2\div 3)$\%.

It is interesting to compare with the data 
the theoretical predictions for the electron azimuthal anisotropy $v_2$,
which is sensitive to the $L$-dependence of the heavy quark energy loss. 
In Fig.~5 we compare our calculations for 
the 20--40\% centrality bin to $v_{2}$ from 
(a) PHENIX \cite{PHENIX2_e} and (b) ALICE \cite{ALICE_e}. 
One sees that the agreement
with the ALICE data is fairly good. However, the experimental errors
are very large and the $p_{T}$ range is too limited 
to make a definitive conclusion on the preferred value of $\alpha_{s}^{fr}$.
For the PHENIX data \cite{PHENIX2_e} $p_{T}\lsim 4$ GeV. For such
low $p_{T}$  our calculations are not robust. Nevertheless, our
$v_2$ for $\alpha_{s}^{fr}=0.5$ (favored by data on 
$R_{AA}$ for pions) at $p_{T}=6$ GeV matches reasonably well the experimental
$v_{2}$ at $p_{T}\approx 4$ GeV.

\begin{figure} [t]
\begin{center}
\epsfig{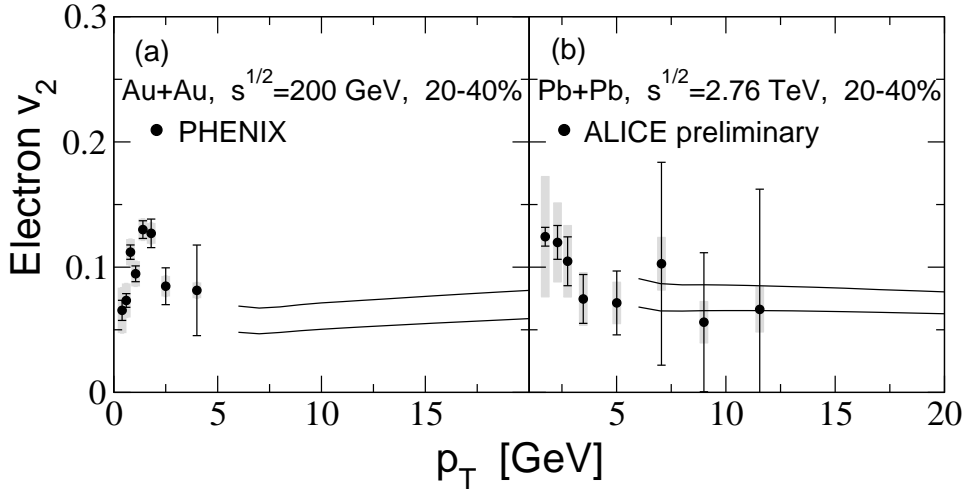}
\end{center}
\caption[.]
{
$v_2$ for single electrons 
for 20--40\% centrality class
in (a) 
Au+Au collisions at $\sqrt{s}=200$ GeV, and in (b) 
Pb+Pb collisions at $\sqrt{s}=2.76$ TeV. 
The theoretical curves are for 
$\alpha_{s}^{fr}=0.4$ (lower curves), and 0.5 (upper curves),
and include
both the charm and bottom contributions.
Data points are from (a) PHENIX \cite{PHENIX2_e}, and 
(b) ALICE \cite{ALICE_e}.
Systematic errors
are shown as shaded areas. 
}
\end{figure}

Note that our predictions were obtained with the radiative energy loss
for the QGP modelled by a system of the static Debye  screened 
color centers \cite{GW}.
The generalization to the dynamical QGP described within the HTL scheme
is trivial (see \cite{AZ_ph} for details). It is reduced to 
replacement of the potential (\ref{eq:a31}) (let us call it $v_{stat}$) 
by a dynamical potential 
$v_{dyn}$, which can be expressed through the gluon polarization operator.  
In the HTL scheme an elegant formula derived in \cite{AGZ}
allows to write $v_{dyn}$ similarly to the static case 
just replacing the factor $1/(q^{2}+\mu_{D})^{2}$
by $1/[q^{2}(q^{2}+\mu_{D})]$ in the formula for the dipole
cross section (\ref{eq:a50}), and increasing a little the overall 
normalization by a factor 
$\frac{\pi^{2}}{6\cdot 1.202}(1+N_{f}/6)/(1+N_{f}/4)\approx 1.19$ (for $N_{f}=2.5$).
This modification leads to unlimited growth of $v_{dyn}$ at large
$\rho$ (due to zero magnetic mass in the HTL approximation), 
while $v_{stat}$ flattens at $\rho\gsim 1/m_{D}$.
It has been recently claimed \cite{MD2} that the dynamical effects
enhance the heavy flavor suppression, and are important for 
description of the single electron suppression. 
Of course, modification of the potential in the above manner
for the dynamical QGP should enhance the heavy quark suppression.
However, this enhancement itself is uninteresting in the context
of the heavy-to-light ratio $R_{AA}^{heavy}/R_{AA}^{light}$ since 
the dynamical potential should enhance the light flavor suppression 
as well. One can expect that the dynamical effects 
affect the induced gluon emission for light and heavy flavors 
similarly.
Indeed, 
the dominating $\rho$-region for induced gluon emission
for light and heavy flavors\footnote{For heavy quarks the typical
  $\rho$ becomes smaller than for light partons only in the tail region
$x\gsim m_{g}/m_{Q}$, where quark mass suppresses strongly
gluon emission.}, is $\rho\lsim \sqrt{S_{LPM}}/m_{g}$
\cite{LCPI} (here $S_{LPM}$ is the suppression factor due to the 
Landau-Pomeranchuk-Migdal
effect, which is typically not very small ($\sim 0.2 - 0.5$) 
for RHIC and LHC).
In this $\rho$-region the shapes of $v_{dyn}(\rho)$ and $v_{stat}(\rho)$ are
qualitatively very similar. For this reason modification of the
nuclear suppression due to the dynamical effects should be similar
for light and heavy flavors. And, once the coupling is fixed by the data on 
$R_{AA}$ for light
hadrons, the dynamical formulation should give $R_{AA}$
for heavy flavors (and single electrons) close to that for
the static model. We have checked this performing numerical calculations
(for a fixed coupling constant as in the HTL scheme), and have found
a negligible modification of the heavy-to-light ratio for the dynamical
model. 
Note that the use of the $v_{dyn}(\rho)$ hardly makes the calculations 
more robust.
Indeed, 
the difference between $v_{dyn}(\rho)$ and $v_{stat}(\rho)$
at $\rho\gsim 1/m_{D}$ is related mostly to the zero magnetic mass in the HTL
scheme. But the lattice calculations 
\cite{magnetic_Md} show that in reality the magnetic mass may be of 
the order of the electric Debye mass. On the other hand, in
the region $\rho\ll 1/m_{D}$, where the HTL approximation is not supposed to be
valid, it gives incorrect normalization of the potential
(contrary to the static model, which gives the correct result
at $\rho\to 0$).

\section{Summary}
In this paper we have examined  whether it is possible
in the pQCD picture of the parton
energy loss 
to simultaneously   describe  experimental 
data from RHIC and LHC on nuclear suppression of light hadrons and
single electrons from the heavy meson decays. 
We have performed calculations taking into account 
radiative and collisional energy loss, 
and fluctuations of the fast parton path lengths in the QGP. 
The calculations of radiative energy loss have been performed
within the LCPI approach \cite{LCPI}.
Both radiative and collisional energy loss were computed 
with running $\alpha_{s}$ frozen at low momenta at
some value $\alpha_{s}^{fr}$. 

We  have found that once the value of 
$\alpha_{s}^{fr}$ is fixed from the data on the nuclear 
modification factor for light
hadrons it gives a satisfactory agreement with the 
data on the electron $R_{AA}$ as well. 
For the electron azimuthal anisotropy $v_2$
the agreement is also within the experimental
errors.
Our calculations show that the effect of collisional
energy loss is realtively small, and cannot be crucial 
for the  flavor dependence of the nuclear suppression factor
for $p_{T}\gsim 10$ GeV.
The collisional mechanism becomes very important only 
for the bottom contribution to the electron spectrum 
at momenta $\lsim 6 - 8$ GeV.

Our results, together with fairly good agreement with
the ALICE LHC data on $R_{AA}$ of $D$-mesons 
\cite{ALICE_RAA_D1,ALICE_RAA_D2} obtained in our recent analysis \cite{RAA12},
give support for the pQCD 
picture of parton energy loss both for light and heavy flavors.
However, the available data on the nuclear suppression
of single electrons, especially from RHIC, analyzed in the present work, 
are restricted to rather low $p_{T}$, where the conditions of applicability
of our pQCD model may be not good enough. 
For more conclusive test of the model it is highly desirable 
to have data on the electron $R_{AA}$ that extend to larger values of $p_{T}$.

Note that measurement of the electron $R_{AA}$ at larger $p_{T}$ 
(say, for $p_{T}\sim 50$ GeV that corresponds to heavy quark momenta
$\sim 100$ GeV) at
LHC is especially desiarble in the light of the
preliminary CMS data \cite{CMS_bjet} on 
the $b$-jet $R_{AA}$ in Pb+Pb collisions at $\sqrt{s}=2.76$ TeV, which
has been measured to be
$0.48\pm 0.09(\mbox{stat.)}\pm 0.18(\mbox{syst.})$ 
at $100<p_{T}<120$ GeV for 0--100\% centrality, while for inclusive jets
$R_{AA}=0.5\pm0.01(\mbox{stat.)}\pm 0.06(\mbox{syst.})$.
This result, if confirmed by 
future more accurate measurements, will be a serious challenge for the
pQCD picture of jet quenching. 
Indeed,
from Fig.~1 one can see that at 
$p_{T}\gsim 100$ GeV the energy losses for light quarks and $b$-quarks
become very close, and their jet $R_{AA}$ should be close as well. 
But at $p_{T}\sim 100 - 120$ GeV
$\sim 60 - 70$\% of inclusive jets are gluon jets that have energy loss
enhanced by a factor $\sim 9/4$, and should have $R_{AA}$
smaller than that for light and heavy quarks. 
For this reason the inclusive jet $R_{AA}$ should be smaller than that
for $b$-jets.

\vspace {.7 cm}
\noindent
{\large\bf Acknowledgements}

\noindent
I am indebted to the referee, who remarked on
the preliminary CMS data \cite{CMS_bjet} on the $b$-jet $R_{AA}$.
This work is supported 
in part by the 
grant RFBR
12-02-00063-a and the program
SS-6501.2010.2.

\renewcommand{\theequation}{A\arabic{equation}}
\setcounter{equation}{0}
\section*{Appendix: Formulas for one gluon $x$-spectrum}
We use the representation of
the one gluon emission $x$-distribution obtained in \cite{Z04_RAA}
which is convenient for numerical calculations. For 
$q\to g q$ process it reads
\beq
\frac{d P}{d
x}=
\int\limits_{0}^{L}\! d z\,
n(z)
\frac{d
\sigma_{eff}^{BH}(x,z)}{dx}\,,
\label{eq:a10}
\eeq
where $n(z)$ is the medium number density, $d\sigma^{BH}_{eff}/dx$ 
is an effective Bethe-Heitler
cross section accounting for both the Landau-Pomeranchuk-Migdal 
and finite-size effects.
The $d\sigma^{BH}_{eff}/dx$
reads 
\beq
\frac{d
\sigma_{eff}^{BH}(x,z)}{dx}=-\frac{P_{q}^{g}(x)}
{\pi M}\mbox{Im}
\int\limits_{0}^{z} d\xi \alpha_{s}(Q^{2}(\xi))
\left.\frac{\partial }{\partial \rho}
\left(\frac{F(\xi,\rho)}{\sqrt{\rho}}\right)
\right|_{\rho=0}\,\,.
\label{eq:a20}
\eeq
Here 
$P_{q}^{g}(x)=C_{F}[1+(1-x)^{2}]/x$ is the usual splitting
function for $q\to g q$ process,
$
M=Ex(1-x)\,
$
is the reduced "Schr\"odinger mass",
$Q^{2}(\xi)=aM/\xi$ with $a\approx 1.85$ \cite{Z_Ecoll},
$F$ is the solution to the radial Schr\"odinger 
equation for the azimuthal quantum number $m=1$ 
\beq
\hspace{-.2cm} i\frac{\partial F(\xi,\rho)}{\partial \xi}=
\left[-\frac{1}{2M}\left(\frac{\partial}{\partial \rho}\right)^{2}
+v(\rho,x,z-\xi)
+\frac{4m^{2}-1}{8M\rho^{2}}
+\frac{1}{L_{f}}
\right]F(\xi,\rho)\,
\label{eq:a30}
\eeq
with the boundary condition
$F(\xi=0,\rho)=\sqrt{\rho}\sigma_{3}(\rho,x,z)
\epsilon K_{1}(\epsilon \rho)$  
($K_{1}$ is the Bessel function),
$L_{f}=2M/\epsilon^{2}$
with $\epsilon^{2}=m_{q}^{2}x^{2}+m_{g}^{2}(1-x)^{2}$,
$\sigma_{3}(\rho,x,z)$ is the cross section of interaction
of the $q\bar{q}g$ system with a medium constituent
located at $z$.
The potential $v$ in (\ref{eq:a30}) reads
\beq
v(\rho,x,z)=-i\frac{n(z)\sigma_{3}(\rho,x,z)}{2}\,.
\label{eq:a31}
\eeq
The $\sigma_{3}$ is given by
\cite{NZ_sigma3}
\beq
\sigma_{3}(\rho,x,z)=\frac{9}{8}
[\sigma_{q\bar{q}}(\rho,z)+
\sigma_{q\bar{q}}((1-x)\rho,z)]-
\frac{1}{8}\sigma_{q\bar{q}}(x\rho,z)\,,
\label{eq:a40}
\eeq
where
\beq
\sigma_{q\bar{q}}(\rho,z)=C_{T}C_{F}\int d\qb
\alpha_{s}^{2}(q^{2})
\frac{[1-\exp(i\qb\ro)]}{[q^{2}+\mu^{2}_{D}(z)]^{2}}\,
\label{eq:a50}
\eeq
is the local  dipole cross section for the color singlet $q\bar{q}$ pair
($C_{F,T}$ are the color Casimir for the quark and thermal parton 
(quark or gluon), $\mu_{D}$ is the local Debye mass).

For $g\to gg$ one should replace the splitting function and $m_{q}$
by $m_{g}$ in $\epsilon^{2}$. The $\sigma_{3}$ in this case reads
\beq
\sigma_{3}(\rho,x,z)=\frac{9}{8}
[\sigma_{q\bar{q}}(\rho,z)+
\sigma_{q\bar{q}}((1-x)\rho,z)
+\sigma_{q\bar{q}}(x\rho,z)]\,.
\label{eq:a60}
\eeq
 
\vskip .5 true cm

\end{document}